\documentclass[twoside,leqno,twocolumn]{article}
\usepackage[letterpaper]{geometry}
\usepackage{ltexpprt}
\usepackage{hyperref}
\usepackage{graphicx}
\usepackage{caption}
\usepackage{subcaption}

\begin{document}

\title{Very Large Language Model as a Unified Methodology of Text Mining}
\author{Meng Jiang\thanks{Department of Computer Science and Engineering, University of Notre Dame, Indiana 46556, USA. Email: mjiang2@nd.edu}}

\date{}

\maketitle

\fancyfoot[R]{\scriptsize{Copyright \textcopyright\ 2023 by SIAM\\
Unauthorized reproduction of this article is prohibited}}

\begin{abstract} \small\baselineskip=9pt
Text data mining is the process of deriving essential information from language text. Typical text mining tasks include text categorization, text clustering, topic modeling, information extraction, and text summarization. Various data sets are collected and various algorithms are designed for the different types of tasks. In this paper, I present a blue sky idea that very large language model (VLLM) will become an effective unified methodology of text mining. I discuss at least three advantages of this new methodology against conventional methods. Finally I discuss the challenges in the design and development of VLLM techniques for text mining.

\end{abstract}

\section{Introduction}
\begin{figure}[t]
    \centering
    \begin{subfigure}[b]{0.48\textwidth}
         \centering
         \includegraphics[width=\textwidth]{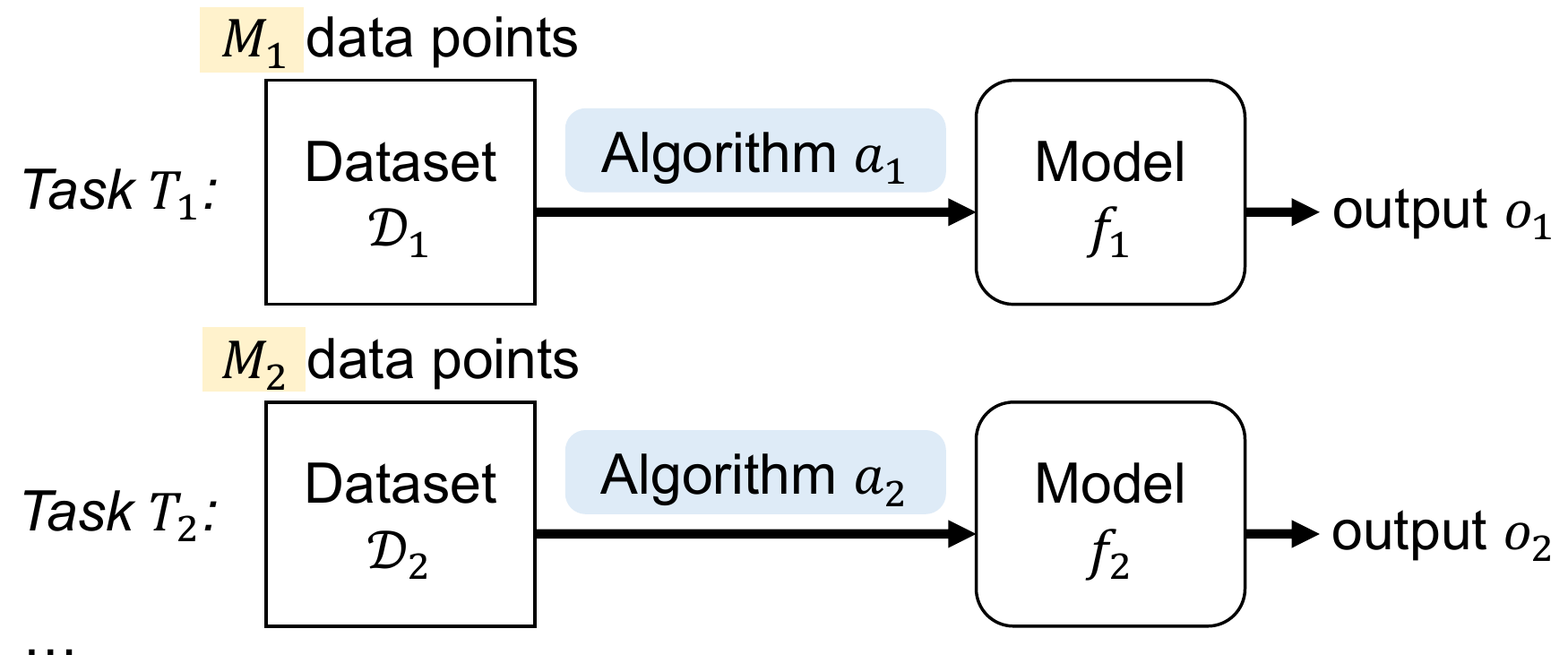}
         \caption{Conventional methodology}
         \label{fig:intro1}
     \end{subfigure}
     \begin{subfigure}[b]{0.48\textwidth}
         \centering
         \includegraphics[width=\textwidth]{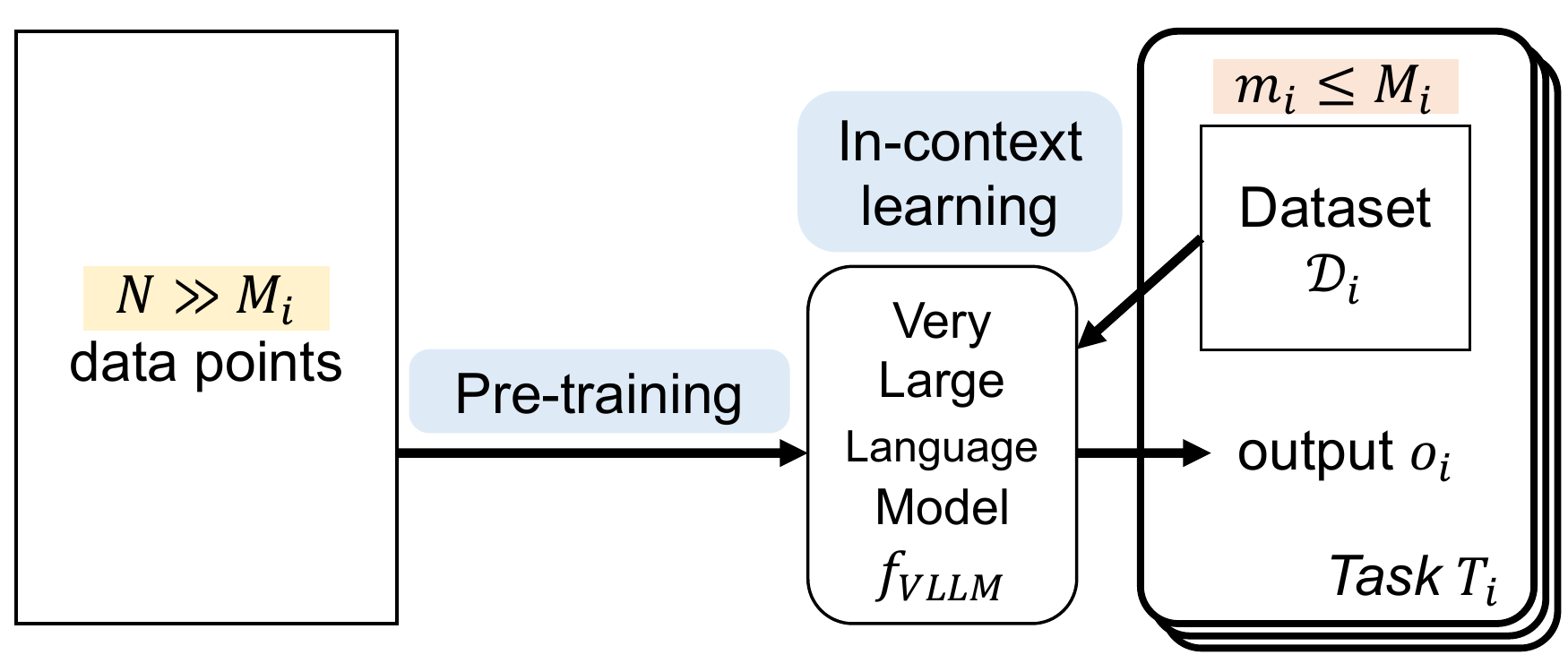}
         \caption{VLLM as a unified methodology}
         \label{fig:intro2}
     \end{subfigure}
    \caption{Compared against conventional methodology, VLLM has three advantages in text mining: (1) unifying tasks with no need of designing specialized algorithms (in blue), (2) knowledge from a much larger size of text data and model (in yellow), and (3) ability of performing a task with a small number of data points (in red).}
    \label{fig:intro}
    \vspace{-0.03in}
\end{figure}

Text data mining is to derive high-quality information from unstructured data. It enables decision-makers to analyze massive amounts of text quickly.
For example, given one million restaurant reviews, one would look for automated solutions to categorizing each piece of review, identifying the aspects it describes, grouping the reviews, detecting topics, and making short summaries.
In data mining and natural language processing (NLP), these are referred as text mining tasks, such as text classification, information extraction, text clustering, topic modeling, and text summarization. For each task, there are communities of researchers who are dedicated to design specialized algorithms and develop models.

\begin{figure*}
    \centering
    \includegraphics[width=0.78\linewidth]{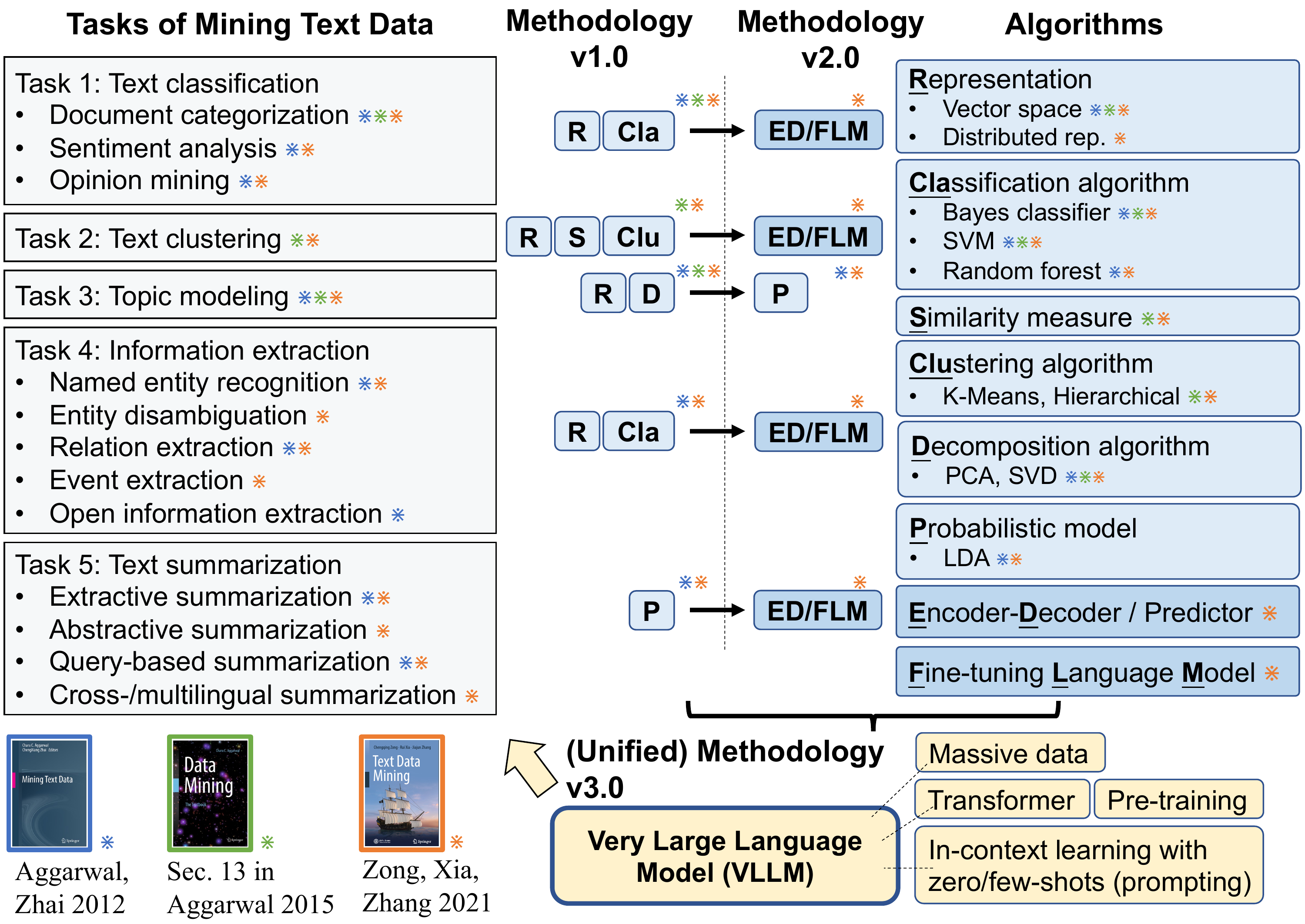}
    \caption{Three influential textbooks on mining text data \cite{aggarwal2012mining,aggarwal2015mining,zong2021text} summarize the algorithms and methodologies on five tasks. Very large language model will become an effective, unified methodology for all the tasks. (Blue, green, and orange stars denote what were covered in the corresponding textbooks.)}
    \label{fig:review}
\end{figure*}

Figure~\ref{fig:intro1} presents the conventional methodology of text mining where one model $f_i$ is developed for one task $T_i$. There are at least two factors on the model's effectiveness: a data set of numerous quality data points $\mathcal{D}_i$ and a properly designed algorithm $a_i$. Deep learning is replacing traditional mining algorithms for most of the tasks. This trend originated from the architecture of neural encoder-decoder/predictor
\cite{xu2017variational,cao2015novel,lample2016neural}, and when the pre-trained bidirectional transformer encoders emerged (known as BERT \cite{kenton2019bert}), fine-tuning the model parameters with $\mathcal{D}_i$ becomes the algorithm that builds a dedicated model $f_i$ on task $T_i$.

Today very large language models (VLLMs) such as GPT-3 family (e.g., \textit{davinci-003}) and ChatGPT are such powerful \textit{text generation} models that they have revolutionized many NLP use cases. They are decoder only unidirectional autoregressive models; they have billions of parameters (much bigger than BERT); they are pre-trained on Internet-scale data and supervised using reinforcement learning from human feedback. Different from BERT, VLLM attempt to replace the downstream fine-tuning with few-shot learning.
They have demonstrated extraordinary learning abilities on story writing, question answering, or mathematical reasoning just by conditioning on input-output examples in form of textual ``prompts'', without optimizing any parameters. VLLMs can possibly perform the text mining tasks as long as their users find proper examples.

In fact, I am presenting a blue sky idea in this paper: VLLM will become a dominant approach of text mining -- an effective unified methodology for various tasks.
Figure~\ref{fig:intro2} shows the data sources, algorithms, and training procedures of VLLMs for text mining. There are at least three advantages of using the VLLMs than conventional methods (such as fine-tuning BERT).
Meanwhile, due to the huge model size, very few people (especially in academia) have resources to fine-tune the VLLMs. So the conventional training diagram would not work with the VLLMs and there are at least three challenges in practice. Next sections will discuss the advantages, approach details, and challenges.

\section{Significance}
VLLM will become the new text mining methodology in next years. Figure~\ref{fig:intro2} briefly presents how VLLMs could be used to perform various tasks. First of all, VLLMs are trained on a dataset of $N$ tokens (much larger than the size of any text mining dataset $M_i$) to predict the next token given the preceding text. This simple objective paired with the large dataset and model results in a very flexible LM that can ``read'' any text input and condition on it to ``write'' text that could plausibly come after the input.

In task $T_i$, the text input can be a concatenation of $m_i$ textual examples selected from $\mathcal{D}_i$ of $M_i$ data points. If the task is text classification, given a new textual example $X'$, suppose we select $\{(X_j, Y_j)\}_{j=1}^{m_i}$ where $m_i$ is usually set as $3$ considering the model's limit of text input length, and $Y_j$ denotes the category label of the $j$-th textual example $X_j$. The text input is:
\begin{quote}
$X_1$ Category : $Y_1$ . $X_2$ Category : $Y_2$ . $X_3$ Category : $Y_3$ . $X'$ Category :
\end{quote}
We expect the next token(s) generated by VLLM to be the predicted category $Y'$. The procedures are similar for other tasks such as topic modeling (i.e., generate a topic's word or special token) and information extraction (i.e., generate entities, their types, and relations).

This new methodology is significantly different from two conventional methodologies of text mining (see Figure~\ref{fig:review}). In ``v1.0'', the first step was extracting representations with vector space model (VSM) or distributed representation learning; then classification, clustering, or other types of algorithms were designed upon the representations for target tasks. In ``v2.0'', neural encoders or pre-trained transformer encoders were employed to automatically extract the representations. Three classical textbooks \cite{aggarwal2012mining,aggarwal2015mining,zong2021text} introduced these two methodologies in which the representations were extracted from moderate sized data and/or models.

Regarding how representations are extracted and how they are used for text mining, the VLLM methodology will have at least three advantages:

\textbf{One model and one algorithm, many tasks:} VLLM is a deep decoder-only model of a huge number of parameters. It extracts representations in the decoder's transformer layers. The model's depth and scale make the representation extraction more flexible than preprocessing tools such as VSM or neural encoders, adapting to the task-oriented text input-output pairs.
One would no longer have to design different specialized algorithms for different mining tasks, which is an expensive and exhaustive job. Instead, one would just need one single VLLM and one decoding algorithm to learn and use the representations to generate desired output.

\textbf{Notably larger data, better representations:} BERT was trained on 2.5B tokens Wikipedia and 800M tokens BookCorpus. In contrast, GPT-3 was trained on 429B tokens WebText, 3B tokens Wikipedia, and 68B tokens BookCorpus. Overall, that's 150 times larger pre-training data, generating deep representations from unprecedented generalization and leading to promising performance in many NLP task settings.

\textbf{Small text data can be mined:} Text mining algorithms used to require a big number of data points, like hundreds of thousands, however, in real applications (e.g., submarine technical reports) we usually have just hundreds \emph{or} thousands of data points to do clustering or topic modeling. Conventional methods could not understand the textual examples effectively from infrequent tokens, so they could hardly find interesting clusters or topics. The pre-trained knowledge inside VLLMs will enable the text understanding of rare words or phrases.

\section{Approach}
The current assumption of text mining method development is to train or fine-tune one model for each task by \emph{optimizing the model's parameters} as long as computational resources can support.
It has become a bottleneck of model size since VLLMs emerged and fine-tuning them is an impossible job for almost all researchers.
To break the bottleneck, I suggest the community to design, develop, and evaluate new methods based on the VLLMs to learn from the task dataset and perform various text mining tasks as generating textual outputs \emph{without optimizing hundreds of billions of parameters}.

\subsection{In-Context Learning}

In-context learning was popularized in GPT-3 as a way to use a VLLM to learn tasks given only a few examples. Users give the VLLM a prompt that consists of a list of input-output pairs that demonstrate a task. A test input is appended at the end of the prompt. The VLLM makes a prediction just by conditioning on the prompt and predicting the next tokens. To correctly generate the test output, the model needs to read the training examples to figure out the input distribution, output distribution, input-output mapping, and the formatting.

\subsection{Text Mining Tasks as Text Generation}

Because VLLMs are text-to-text generation models \cite{yu2022survey}, a typical approach is to embed the description of the task in the text input, e.g., as a question instead of it being implicitly given. This is known as \emph{prompt learning} or \emph{prompt engineering}. In few-shot NLP tasks, prompts encourage a chain of thought for multi-step reasoning; in DALL-E and Stable Diffusion, prompts are designed to turn text into images. Towards mining a number of text data examples, the prompts should have not only the task description but also a (sub)set of the data examples to adapt the model to the target mining dataset. Novel prompting techniques are needed on VLLMs to turn text mining into text-to-text generation tasks.

Text categorization can be naturally transformed into prompting with few-shots as given in Section 2. The goal was to accurately distinguish data examples of different class labels. Therefore, if possible, re-defining the category \emph{names} might help the model approach the optimal classification performance. In text clustering, the model may not generate {cluster ID}. One potential idea is to ``ask'' the model how similar two examples are, so any clustering algorithms (e.g., spetral clustering) on the proximity matrix can identify the clusters. Next, topic taxonomies will play a much more significant role in VLLM-based topic modeling, because the model can directly generate the topic names while the traditional techniques (e.g. PLSI, LDA) cannot. For example, in submarine technical reports, the topic names include ``underwater detection capabilities'' and ``silent propulsion system''; it is important to prompt with the taxonomic information and topic-relevant examples.

\section{Challenges}
This section discusses four critical challenges in applying or developing VLLMs for text mining tasks.

\subsection{VLLM with Many Shots}

According to the OpenAI API, requests to VLLMs can use up to 4,097 tokens ($\sim$3,000 words, $n_{1}$=$150$ to $200$ sentences) that are shared between prompt and generation. So the models perform few-shot learning which sounds powerful but limits the learning capacity when many shots are available in text mining. For example, in text clustering if each data example has $n_{2}$=$8$ to $10$ sentences, the model's text input has $14$ to $24$ data examples ($n_{1}/n_{2}$-1). More multi-layer attentions would be needed to process more tokens of text input, in order to learn from more examples. It limits VLLMs to expand learning sets without optimizing the parameters.

\vspace{-0.1in}
\subsection{VLLM with Diverse Prompts / Examples}

Given the constraint of a VLLM's text input length, we cannot use many example or very long prompts, but we can query the model multiple times for each test instance and aggregate the outputs, if time allows. In that case, a good strategy is to try diverse data examples and prompts in each time's input. The model will use \emph{multi-view knowledge} with the diverse inputs to make inferences. Yu et al. found that language models could be augmented by retrieving textual inputs from multiple knowledge sources \cite{yu2022retrieval}. They also discovered that using diversified prompts (i.e., clustering prompts and selecting representative ones from each cluster), VLLMs could generate more accurate contexts for question answering (QA) readers than using purely the top prompts similar with questions \cite{yu2022generate}. The exciting results in QA indicate that the diverse prompts or data examples can be helpful for the VLLMs on text mining tasks.

\vspace{-0.1in}
\subsection{VLLM with Lifelong Memory}

We can hardly update the huge number of parameters in VLLMs (i.e., fine-tuning) with any updated information. A memory contains a large collection of text elements (e.g., entities, concepts, noun phrases, relational phrases) and their corresponding representation vectors. Zhang et al. developed the first language model accompanied with the memory architecture to augment its text-to-text generation performance \cite{zhang2022unified}. This model extracted and used the representations of the elements that were related to text input from the memory. The representations could be efficiently pre-trained or updated with dynamic data. However, none of the existing VLLMs utilized a memory, and so far none of the memories used the \emph{lifelong machine learning} paradigm that learned continuously, accumulated the knowledge learned in previous tasks or data, and used it to help future learning. Developing a VLLM with lifelong memory is still an open problem.

\vspace{-0.1in}
\subsection{VLLM with Structured Data}

VLLMs leverage natural language corpora that are derived from the Web. However, natural language text alone represents a limited coverage of knowledge \cite{yu2022diversifying}. Existence of non-factual information and toxic content in text can eventually cause biases in the models. 
Alternate sources of information are metadata, data tables, taxonomies, ontologies, and knowledge graphs (KGs), which consist of structured data. For example, KGs are more factual than unstructured data because the information is usually extracted from more trusted sources, and post-processing filters; and human editors may ensure inappropriate and incorrect content are removed. So, the models that incorporate them can improve factual accuracy \cite{zhu2020enhancing} and reduce toxicity.



\section{Expected Success and Conclusions}
This paper discussed a blue sky idea that is a revolutionized text mining methodology -- designing and developing new techniques based on very large language model (VLLM). It presented the significance, potential approaches, and challenges of this idea in various tasks or applications. The expected success is to achieve a significantly higher level of text mining performance than the conventional text mining techniques.

\bibliographystyle{unsrt}
\bibliography{main}

\end{document}